\documentclass[a4paper]{jpconf}
\usepackage{graphicx}

\usepackage{hyperref}

\usepackage{amsmath}
\usepackage{amssymb}

\usepackage{subcaption}
\usepackage{longtable}
\usepackage{booktabs}
\usepackage{listings}

\def\tightlist{}
\def\citep{\cite}

\begin{document}

\title{Everware toolkit. Supporting reproducible science and challenge-driven
education.}

\author{A Ustyuzhanin$^{1,2}$, T Head$^{3}$, I Babuschkin$^{4}$, A Tiunov$^{2}$}

\address{$^{1}$National Research University Higher School of Economics}
\address{$^{2}$Yandex School of Data Analysis}
\address{$^{3}$Wild Tree Technologies}
\address{$^{4}$Manchester University}

\ead{andrey.u@gmail.com}

\begin{abstract}
\noindent Modern science clearly demands for a higher level of reproducibility and
collaboration. To make research fully reproducible one has to take care
of several aspects: research protocol description, data access,
environment preservation, workflow pipeline, and analysis script
preservation. Version control systems like git help with the workflow
and analysis scripts part. Virtualization techniques like Docker or
Vagrant can help deal with environments. Jupyter notebooks are a
powerful platform for conducting research in a collaborative manner. We
present project Everware that seamlessly integrates git repository
management systems such as Github or Gitlab, Docker and Jupyter helping
with a) sharing results of real research and b) boosts education
activities. With the help of Everware one can not only share the final
artifacts of research but all the depth of the research process. This
been shown to be extremely helpful during organization of several data
analysis hackathons and machine learning schools. Using Everware
participants could start from an existing solution instead of starting
from scratch. They could start contributing immediately. Everware allows
its users to make use of their own computational resources to run the
workflows they are interested in, which leads to higher scalability of
the toolkit.
\end{abstract}

\newcommand{\plusnamesingular}{}
\newcommand{\starnamesingular}{}
\newcommand{\xrefname}[1]{\protect\renewcommand{\plusnamesingular}{#1}}
\newcommand{\Xrefname}[1]{\protect\renewcommand{\starnamesingular}{#1}}
\providecommand{\cref}{\plusnamesingular~\ref}
\providecommand{\Cref}{\starnamesingular~\ref}
\providecommand{\crefformat}[2]{}
\providecommand{\Crefformat}[2]{}

\crefformat{figure}{fig.~#2#1#3}
\Crefformat{figure}{Figure~#2#1#3}

\section{Introduction}\label{introduction}

The demand for reproducibility in scientific research is gradually
increasing nowadays. It could be seen by number of re-research cases, in
which people are spending considerable amount of efforts to replicate
results obtained by others before. Popular cases of such replication
efforts can be found in biology \citep{prinz2011believe}, psychology
\citep{psycho2015rep} and other branches of the science \citep{lid1500}.
One of the main outcome of such research is that fraction of the results
that coincide with original published result is considerably lower than
expected. ``More than 70\% of researchers have tried and failed to
reproduce another scientist's experiments, and more than half have
failed to reproduce their own experiments'' - those are some of the
telling figures that emerged from Nature's survey of 1,576 researchers
who took a brief online questionnaire on reproducibility in research
\citep{lid1500}.

Another aspect of scientific efforts that is closely related to research
and hence to the reproducibility is so-called ``challenge driven
education''. Examples of that kind of efforts include hackathons,
interactive workshops, seminars, master classes and other kinds of
outreach activities. These hackathons usually suppose that participants
a) not equally well-familiar with the problem they are supposed to
tackle, but b) they should start from same baseline solution that should
be easily accessible and understandable for everyone participating. Also
results participants obtain during the event should also be accessible
and reproducible at least for the organisers and perhaps to all the
participants.

For the sake of technical-level discussion and without touching much of
the social reasons of research irreproducibility, let us consider the
case of computational experiment. It is usually takes great deal of
efforts for every experiment, but it starts as data has been collected
and ends at the point as experimental results firmly confirm/refute on
of the hypothesis under the consideration. Even improving
reproducibility of that part of the experiment may lead to much better
understanding of the particular results by the community. Namely
accessibility and repeatability of the analysis pipelines of successful
researchers can improve the following aspects: better
mentoring/supervision, more careful within-lab validation, simplified
external-lab validation, give incentive for better practice and robust
experiment design. From educational point of view such transparency
would widen access to the best practices and facilitate better teaching.
The aspects mentioned above are highly correlated with the
reproducibility boost factors mentioned in survey \citep{lid1500}.

\section{Reproducibility key
components}\label{reproducibility-key-components}

Let us consider the main aspects of the computational research that lead
to successful reproducibility of the results. It usually starts from
rather basic level of understanding of \emph{assumptions} that are
effective in the field of the research. For example to reproduce results
in the field of high energy physics you have to understand basic
concepts of event, likelihood fit, decay, mother/daughter particles and
so on. As soon as one is comfortable with those definitions and
assumptions, he/she can follow definition of the data and understand
columns/relationships between different entities. Another item included
into assumptions would be description of hypothesis being evaluated.
Accessibility of the \emph{dataset} and its metadata is the second
degree of the reproducibility. Then follows description of the
\emph{computational environment} that should be created/restored for the
experiment to run. Depending on the level of detalization such
description may contain just name of the interpreter and operating
system, or can be as detailed as list of required modules, system
libraries, configuration files along with exact specification of its
versions. Some guidelines demand nothing less than virtual machine with
all the modules pre-installed. Next component to follow is the
\emph{code of the computational analysis} that reads data, performs
necessary transformations and computes final analysis numbers that lead
to certain conclusions about hypothesis being considered within the
analysis. In some cases the analysis performed by the code is split into
various steps: data profiling, feature engineering, simulation, etc.
Such steps may form a certain \emph{workflow} that researcher should be
able to trigger by a single command. Description of the workflow can be
stored in a simple \texttt{Makefile}, or could be more sophisticated.
Some of the best-practices of software engineering \citep{testing}
advise that to stay on the safe side developing complicated piece of
software (which is not uncommon to modern computational experiments),
one should iteratively invest in \emph{automated cross-checks} (which in
software engineering are called unit tests). Such checks are not
something necessary, but might be a good candidate for generation of
intermediate report of the experiment that helps researchers to
understand overall status of the analysis. The final component of the
computational experiment would be the conclusion supported by
\emph{concrete number/figures} that should be generatable by executing
the certain command/workflow to produce the results from the data (or
links to that data).

Just to summarize computational experiment ingredients:

\begin{itemize}
\tightlist
\item
  assumptions;
\item
  data, metadata;
\item
  environment + resources (CPU/GPU);
\item
  code/scripts;about re-usable science, it allows people to jump right
  into your research code. Lets you launch \emph{Jupyter} notebooks from
  a git repository with a click of a button.
\item
  workflows;
\item
  automated intermediate results checks;
\item
  final results (datasets, figures).
\end{itemize}

\section{Ideal-world picture}\label{ideal-world-picture}

So the overall picture of the reproducible analysis would include the
following stages:

\begin{itemize}
\tightlist
\item
  preparation of the environment, specification;
\item
  preparation of the dataset for the analysis;
\item
  preparation of the software repository for storing the code-related
  items;
\item
  iterative writing/testing analysis code; updating environment
  specification if necessary;
\item
  automating running of complicated workflows, e.g.~as
  \texttt{Makefile};
\item
  automating checking of the intermediate results;
\item
  as scientific results has been confirmed by the code, one can publish
  along the paper:
\item
  the repository with the specification, code, workflow, etc;
\item
  the dataset (or some derivative that is required to finish some of the
  final steps);
\item
  if needed, it is also possible to publish environment snapshot;
\item
  so other members of scientific community interested in the
  following-up the research can bind repository, datasets and
  environment together.
\end{itemize}

As the final set of publishing steps is completed, and one is looking
for reproduction of the results, play with the code, etc. Usually the
latter step involves a lot of hassle but it is a good place for Everware
to come in. Everware (\url{https://github.com/everware/everware}) is
about re-usable science, it allows people to jump right into your
research code. Lets you launch your analysis code from a git repository
with a click of a button. Namely, Everware can read environment
specification from the code repository, create the environment, put all
the code/workflow inside this environment and allow user to interact
with it. Of course, some additional assumptions should be taken into
account and we are describing it in the following sections.

\section{Everware-compatible
repository}\label{everware-compatible-repository}

The suggested approach for storing components mentioned in the previous
section heavily relies on modern version control system, namely git.
Indeed, configuration of the research components can be stored in the
form of text files that easily fit into git storage system with all
powerful features git provides.

\begin{longtable}[]{@{}ll@{}}
\caption{Computational research components that could be stored in git
repository.}\tabularnewline
\toprule
\begin{minipage}[b]{0.41\columnwidth}\raggedright\strut
\textbf{component}\strut
\end{minipage} & \begin{minipage}[b]{0.53\columnwidth}\raggedright\strut
\textbf{repository entity}\strut
\end{minipage}\tabularnewline
\midrule
\endfirsthead
\toprule
\begin{minipage}[b]{0.41\columnwidth}\raggedright\strut
\textbf{component}\strut
\end{minipage} & \begin{minipage}[b]{0.53\columnwidth}\raggedright\strut
\textbf{repository entity}\strut
\end{minipage}\tabularnewline
\midrule
\endhead
\begin{minipage}[t]{0.41\columnwidth}\raggedright\strut
assumptions\strut
\end{minipage} & \begin{minipage}[t]{0.53\columnwidth}\raggedright\strut
\texttt{README.md} (and other documentation media)\strut
\end{minipage}\tabularnewline
\begin{minipage}[t]{0.41\columnwidth}\raggedright\strut
environment\strut
\end{minipage} & \begin{minipage}[t]{0.53\columnwidth}\raggedright\strut
\texttt{Dockerfile}\strut
\end{minipage}\tabularnewline
\begin{minipage}[t]{0.41\columnwidth}\raggedright\strut
code/scripts\strut
\end{minipage} & \begin{minipage}[t]{0.53\columnwidth}\raggedright\strut
source code\strut
\end{minipage}\tabularnewline
\begin{minipage}[t]{0.41\columnwidth}\raggedright\strut
workflows\strut
\end{minipage} & \begin{minipage}[t]{0.53\columnwidth}\raggedright\strut
\texttt{Makefile}, \texttt{snakefile}, etc\strut
\end{minipage}\tabularnewline
\begin{minipage}[t]{0.41\columnwidth}\raggedright\strut
automated intermediate results check\strut
\end{minipage} & \begin{minipage}[t]{0.53\columnwidth}\raggedright\strut
\texttt{circle.yml} (continuous integration config file)\strut
\end{minipage}\tabularnewline
\begin{minipage}[t]{0.41\columnwidth}\raggedright\strut
final results\strut
\end{minipage} & \begin{minipage}[t]{0.53\columnwidth}\raggedright\strut
\texttt{html}, \texttt{pdf}, \texttt{png}, etc.\strut
\end{minipage}\tabularnewline
\bottomrule
\end{longtable}

One specific approach we find quite comfortable for storing environment
configurations relies on Docker \citep{docker}. By means of using
virtualization capabilities of modern operating systems, Docker allows
for creation of a virtual machine in very light-weight manner compared
to traditional approaches. \texttt{Dockerfile} is essentially the set of
instructions for creating environment that the code should be able to
run at. It has been shown by Boettiger \citep{dockerBoettiger14} that
Docker has enough capabilities for purposes of computational research
reproducibility: it supports versioning of the environment, it is quite
easy to run on local machine (notebook) or on remote server, also you
can share images of your environment via centralized docker image
storage (docker hub).

The source code of the research can be stored in variety of ways. But
there is one quite prominent approach for doing research in interactive
way. Initially the project was called IPython notebooks
\citep{ipython2007}, now it is transformed into multi-language system
called Jupyter \citep{jupyter}. Such notebooks contain mixture of the
source code, rich comments and output in the form of text, tables,
figures or even animation. To execute such notebook one has to start
Jupyter server and connect to it through the web browser. The
environment in which server starts should have all the libraries
installed for the code to run. The user interacting with the notebook
can re-evaluate all the code and theoretically can get the same output.
Jupyter ecosystems nowadays becomes increasingly poplar since it
includes plenty of tools that make analysis review considerably simplier
to follow (think of it as a shift from the regular to \emph{literate
programming}\citep{knuth1984literate}) and it has quite vibrant
community to support all those tools.

Jupyter notebooks (or simply `notebooks') can be executed in batch mode
by means of Jupyter nbconvert \citep{nbconvert} or 3-rd party modules
like runipy \citep{runipy}. So for the cases when research workflow
consists of several steps, it would be quite natural to store commands
of those steps either in \texttt{Makefile}-like file or in simple
\texttt{.sh} script file. Such practice might simplify testing of the
results through the development of the research.

So one of the possible way of structuring computational research
repository would be the following:

\begin{itemize}
\tightlist
\item
  README.md with general information and references;
\item
  \texttt{Dockerfile} that contains instructions for the environment
  configuration;
\item
  Jupyter notebooks with all the code;
\item
  \texttt{Makefile} (or \texttt{.sh} script) to rule them overall
  workflow (optional);
\item
  \texttt{circle.yml} with instructions to run the workflow at CI
  systems (optional).
\end{itemize}

There has been tremendous amount of efforts spent on advocating writing
tests for software engineering\citep{testing}. Such tests basically help
to trace progress of the development and help to monitor the quality of
the software. Continuous integration (CI) systems is the de-facto
standard for execution of the tests. Those systems could be hosted
on-site or publicly available. Usually they differ by capabilities
(i.e.~support for running code under different operating systems or
services support (like Docker)) and configuration approach, but
essentially they are quite similar. Examples of the CI system with great
capabilities and straightforward configuration include CircleIC
\citep{circleci} and Travis. So borrowing those best practices from
software engineering world to scientific research could help . Although
it is not strictly necessary for Everware to proceed.

Here are several examples of the repositories that are structured this
way: \url{https://github.com/openml/study_example},
\url{https://github.com/yandexdataschool/comet-example-ci},
\url{https://github.com/yandexdataschool/neurohack_dec_2016}.

The latter repository is a preparation for a data analysis hackathon on
neural sciences. As mentioned before requirements for reproducibility of
an experimental research are similar to requirements for an educational
challenge. Indeed a) participants have to bootstrap themselves into
problems given, so basic assumptions has to be specified in some way, b)
due to limited amount of time for the challenge, participants have to
start with already prepared environment, c) source code has to be
executable, understandable and changeable, d) workflows and automated
test may help participants to understand complications and constraints
of the challenge. On the following picture there is a flow of actions
for preparation of the repository for publishing a research (a) and
steps required by publisher/organizer of educational challenge and its
participants (b).

\begin{figure}[htbp]
\centering
\includegraphics[width=1.00000\textwidth]{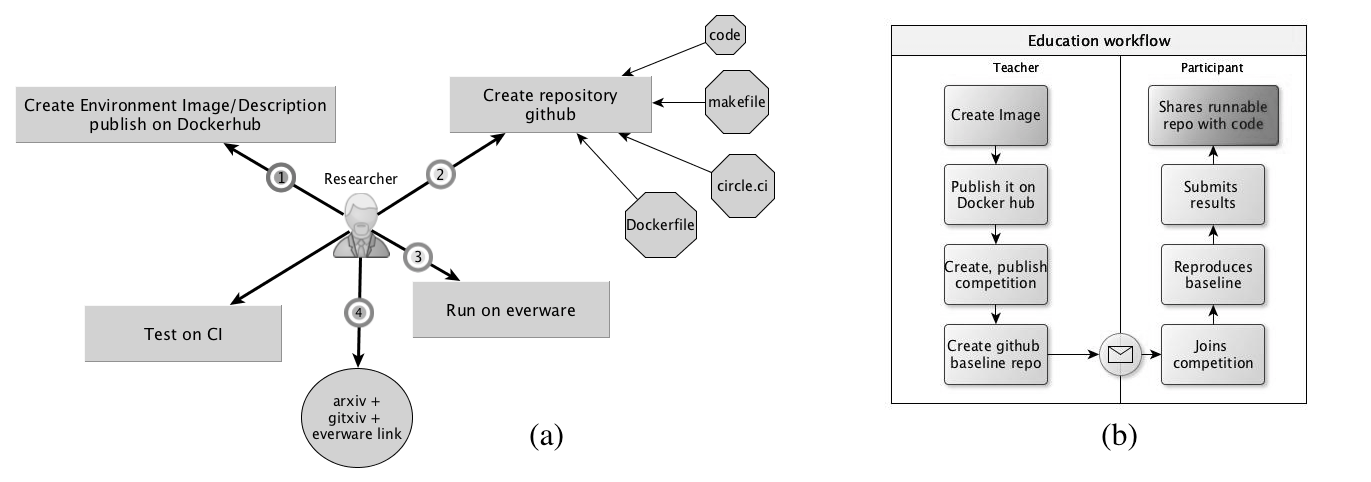}
\caption{Repository preparation for a) research publishing b)
educational challenge\label{fig:1}}
\end{figure}

Those steps are pretty well aligned with ideal-world picture presented
in the section above, although impose some practical restrictions we'll
discuss in the following section.

\section{Architecture}\label{architecture}

\begin{figure}[htbp]
\centering
\includegraphics[width=0.90000\textwidth]{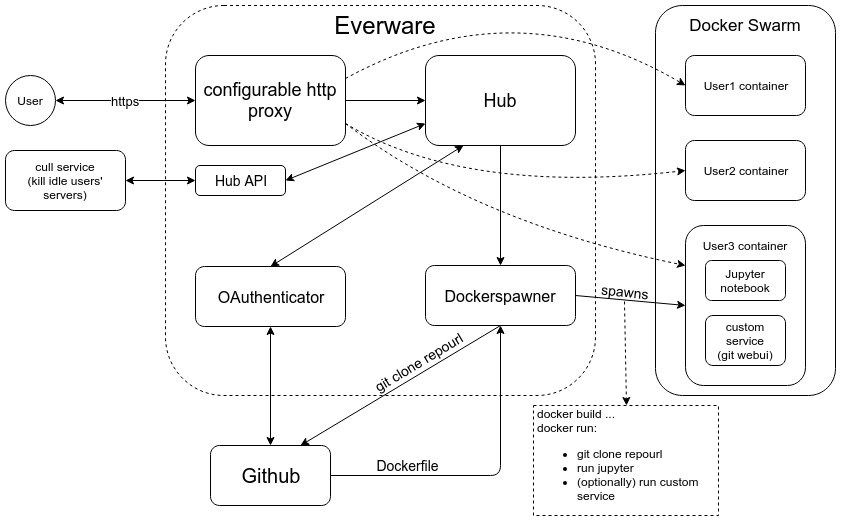}
\caption{Everware architecture.\label{fig:2}}
\end{figure}

Everware is based on JupyterHub\citep{jupyterhub} component. JupyterHub
is a multi-user server that manages and proxies multiple instances of
the single-user Jupyter notebook server. Actually Everware enhances
Jupyterhub with git authentication and repository cloning functionality.
Also Everware enhances spawner component such that it creates custom
Docker image from the \texttt{Dockerfile} in the repository. Upon
successful image creation it is spawned on available Docker cluster.
Overall picture is presented on the \xrefname{fig.}\cref{fig:2}.

Such architecture is pretty simplistic and at the same time flexible
enough to support for extensions we'll talk about in the next section.

\section{Discussion}\label{discussion}

Everware provides a powerful mechanism to support reproducibility and to
our knowledge it has been used for several research environments and
educational challenge setups. As a demonstration stand there is a
computational cluster donated by Yandex
(\url{https://everware.rep.school.yandex.net}). It has been used rather
extensively for educational and research activities:

\begin{itemize}
\tightlist
\item
  Machine Learning in High Energy Physics summer school 2016
  (\href{https://github.com/yandexdataschool/mlhep2016}{github link});
\item
  YSDA course on Machine learning at Imperial College London 2016
  (\href{https://github.com/yandexdataschool/MLatImperial2016}{github
  link});
\item
  ``Flavour of physics'' Kaggle competition, organized in 2015
  (\href{https://github.com/yandexdataschool/flavours-of-physics-start}{github
  link});
\item
  Machine Learning course at University of Eindhoven
  (\href{https://github.com/openml/course-python}{github link});
\item
  LHCb open data masterclass
  (\href{https://github.com/lhcb/opendata-project}{github link});
\item
  YSDA course on Machine Learning at Imperial College London 2017
  (\href{https://github.com/yandexdataschool/MLatImperial2017}{github
  link});
\item
  Hachathon on Neural Sciences organized by Yandex, HSE
  (\href{https://github.com/yandexdataschool/neurohack-2016-starterkit}{github
  link});
\item
  Electro-magnetic showers detection
  (\href{https://github.com/FilatovArtm/opera_project/}{github link}),
  meta-machine learning research
  (\href{https://github.com/openml/Study-14}{github link}),
  reinforcement learning
  (\href{https://github.com/gamers5a/DeepReinforcement}{github link}).
\end{itemize}

Based on our experience such approach would stimulate open research by
the following aspects:

\begin{itemize}
\tightlist
\item
  make easier supervision/mentoring;
\item
  make easier within-lab validation;
\item
  widen access to the best practices;
\item
  simplify cross-lab validation;
\item
  provide good incentive for formal reproduction;
\item
  is a good thing for industry career track development.
\end{itemize}

It was mentioned before that those aspects are pretty well aligned with
reproducibility boosting factors mentioned in Nature survey
\citep{lid1500}. As a cost one would have to pay for those benefits we
see the following burden:

\begin{itemize}
\tightlist
\item
  learning a bit of (open-sourced) technology;
\item
  re-organize internal research process;
\item
  inner barrier for openness;
\item
  higher incentive for mindless \emph{borrowing};
\item
  divergence/potential learning curves (promotes users to create unique
  environments).
\end{itemize}

\subsection{Open questions, project
roadmap}\label{open-questions-project-roadmap}

Current limitations of the proposed approach are:

\begin{itemize}
\tightlist
\item
  dependence on Jupyter computational model. This nowadays may be a
  rather strict suggestion;
\item
  no access to private data sources;
\item
  bottleneck for resource-hungry (either RAM or CPU or disk) analyses -
  i.e.~the system doesn't scales itself with increased demand for
  computationlly-intensive tasks.
\end{itemize}

To address those limitations further development of the project would
include the following aspects:

\begin{itemize}
\tightlist
\item
  Bring Your Own Resources (BYOR) model. This model would allow to
  improve a bit complication with resource-hungry analyses, since user
  would be able to point to his own resources (Amazon, CERN, whatever)
  and instantiate experiment execution there;
\item
  Jupyter kernel inside separate docker container. This would allow for
  easier integration of user repositories since it won't require any
  special things like running Jupyter inside the container and it would
  be much easier to guess required environment configuration;
\item
  support for custom web interface (e.g.~for R shine, or
  git-webui\citep{gitwebui}). This approach would allow for having
  several services accessible to user from the same container and in
  case of git-webui simplify interaction with versioning system a lot;
\item
  storage of time-limited user certificates (or proxies) inside docker
  container during instantiation;
\item
  support automatic capture of the research environment (integration
  with ReproZip\citep{reprozip2016});
\item
  add support for container execution customization, like specification
  of input file sources or additional container(s) that has to be
  started along with the main one.
\end{itemize}

\section{Conclusion}\label{conclusion}

In this paper we presented open-source Everware project that is aimed at
helping in reproducibility of scientific research and educational
activities. The main scope of the project is computational part of the
experiment. From the first look the project may seem a bit of a burden
on shoulders of researcher but our own experience shows that this price
is negligible compared to the reward in the long-term. To help one to
get started there are hints and guidelines readily available at:
\url{https://github.com/everware/everware/wiki/Getting-Started}

\section*{References}

\bibliography{bibliography}

\end{document}